\documentclass[preprint2]{aastex}

\def\c2h5oh {\hbox{${\rm C}_2{\rm H}_5{\rm OH}$}} 
\def\tco {\hbox{$^{13}{\rm CO}$}}                 
\def\cdo {\hbox{${\rm C}^{18}{\rm O}$}}          
\def\ct {\rm J$_{Ka,Kc}$=4$_{1,4}$\raw 3$_{0,3}$} 
\def\sc {\rm J$_{Ka,Kc}$=6$_{3,4}$\raw 5$_{2,3}$} 
\def\no {\rm J$_{Ka,Kc}$=9$_{0,9}$\raw 8$_{1,8}$} 
\def\as {\ifmmode {^{\scriptscriptstyle\prime\prime}} 		       
	\else $^{\scriptscriptstyle\prime\prime}$\fi}
\def\asns {\ifmmode {^{\scriptscriptstyle\prime\prime};}               
        \else $^{\scriptscriptstyle\prime\prime};$\fi}
\def\ta*{$T_{\rm A}^{*}$}
\def\tex {$T_{\rm ex}$}

\chardef\isp="10 
\def\i{\'\isp}
\def\gsim {\ifmmode {\buildrel>\over\sim}		               
	\else {\lower.6ex\hbox{$\buildrel>\over\sim$}}\fi}
\def\lsim {\ifmmode {\buildrel<\over\sim}	   	               
	\else {\lower.6ex\hbox{$\buildrel<\over\sim$}}\fi}
\def\am {\ifmmode {^{\scriptscriptstyle\prime}}			       
	\else $^{\scriptscriptstyle\prime}$\fi}
\def\deg {\ifmmode^\circ\else$^\circ$\fi}			       
		
\def\raw {\ifmmode\rightarrow\else$\rightarrow$\fi}	
\def\uc{\rm J=1\raw0} 
\def\du{\rm J=2\raw1}					
\def\td{\rm J=3\raw2}					

\def\hdos {\hbox{${\rm H}_2$}}                    
\def\cmm#1{\ifmmode {\,{\rm cm^{-#1}}\;} 		 
		\else \hbox{$\,${\rm cm$^{\rm -#1}\;$}}\fi}

\def\masmenos{\ifmmode {\pm} \else $\pm$ \fi}		               

\def\kms{\ifmmode {{\rm \;km\;s^{-1}\;}}		    	      
        \else {\hbox{$\,${\rm km$\;$s$^{\rm -1}\;$}}}\fi}

\def\apro{$\sim$}                                   
\def\gt{\ifmmode {>}\else{$>$}\fi}

\def\c2h5oh{\hbox{${\rm C}_2{\rm H}_5{\rm OH}$}}                
\def\E#1{\ifmmode \,10^{#1}\; \else {${\rm\,10^{#1}}\;$}\fi}	       
\def\T#1{\ifmmode {$\times$10^{#1}\;} \else {\hbox{$$\times$10^{#1}\;$}\fi}}

\lefthead{}
\righthead{}

\begin{document}

\title{Large scale grain mantle disruption in the Galactic Center }

\author{J. Mart{\i}n-Pintado\altaffilmark{1}, J. R. Rizzo\altaffilmark{1}, P. 
de Vicente\altaffilmark{1}, N. J. Rodr{\i}guez-Fern\'andez\altaffilmark{1} and 
A. Fuente\altaffilmark{1}}

\altaffiltext{1} {Observatorio Astron\'omico Nacional, Apartado 1143, E-28800 
Alcal\'a de Henares, Spain}

\begin{abstract}
We present observations of \c2h5oh toward molecular clouds in Sgr A, Sgr B2 
and associated with thermal and non-thermal features in the Galactic center. 
\c2h5oh emission in Sgr A and  Sgr B2 is widespread, but not uniform. \c2h5oh 
emission is much weaker or it is not detected in some molecular clouds in 
both complexes, in particular those with radial velocities between 70 and 120 
\kms. While most of the clouds associated with the thermal features do not 
show \c2h5oh emission, that associated with the Non-Thermal Radio Arc shows 
emission.
The fractional abundance  of \c2h5oh in most of the clouds with radial 
velocities between 0 and 70 \kms in Sgr A and Sgr B2 is relatively high, of few 
\E{-8}. The \c2h5oh abundance decreases by more than one order of magnitude 
(\lsim \E{-9}) in the clouds associated with the thermal features. The large 
abundance of \c2h5oh in the gas-phase indicates that \c2h5oh has 
formed in grains and released to gas-phase by shocks in the last \apro \E{5} 
years.
The implications of this finding in the origin of the shocks in the GC is 
briefly discussed.

\end{abstract} 


\section{Introduction}

Ethanol (\c2h5oh) was first detected by \citet{zuckerman75} towards the massive 
star forming region in the Galactic center (GC) Sgr B2. Subsequent 
observations of
this molecule in the interstellar medium have shown that \c2h5oh  is only 
present in the dense (\gsim\E{6}\cmm3) and hot (\gsim 100K) cores associated 
with newly formed massive stars 
\citep{irvine87,millar88,ohishi95,millar95,nummelin98}. In hot cores, the 
abundance
of \c2h5oh in gas-phase ranges from \E{-8} to \E{-9}, and gas-phase chemistry 
cannot
account even for the lowest fractional abundance of this molecule. Therefore, \c2h5oh 
emission is considered one of the best tracers of dust chemistry 
\citep{millar88,charnley92,charnley95}. 
Furthermore, the transient nature of the alcohol chemistry in molecular clouds, 
makes the abundance of \c2h5oh an excellent clock to estimate when this
 molecule was injected to the gas-phase \citep{charnley95}.
 
The GC molecular clouds show high kinetic temperatures (\gsim 80 K) 
\citep{hutte93, rodriguez01} and  widespread large abundance of refractory 
molecules like SiO \citep{martin97}. Shock waves have been invoked to 
explain the heating, the morphology and the relatively large abundance of SiO 
in the GC \citep{martin97}. Shocks are also expected to 
sputter molecules from 
the icy grain mantles increasing the gas-phase abundance of the molecules 
formed in grains. Evidences for the release of molecules from grain mantles in 
selected molecular clouds close to Sgr A comes from the detection of \c2h5oh 
and HOCO$^+$\ in this region \citep{minh92, charnley00}. In the scenario that shocks drive the
chemistry in the GC clouds, it is expected that widespread 
\c2h5oh emission closely follows that of SiO. 

In 
this letter we present observations of \c2h5oh toward molecular clouds in 
Sgr A, Sgr B2 complexes and associated with the thermal features (Arched 
Filaments --hereafter TAF-- and the 
Sickle) and the Non Thermal Radio Arc (NTRA) in the GC 
\citep[for the nomenclature]{lang99}. We find that the \c2h5oh emission in 
the GC is widespread, but large variations in the abundance of this molecule are 
found between the different GC molecular clouds. The large \c2h5oh 
abundance and its variations are consistent with a scenario 
in which recent (\lsim \E{5} years) shocks dominate the chemistry in the GC
by grain processing.

\section {Observations and results}

The observations of the \ct, \sc\ and \no\ lines of \c2h5oh towards the GC 
were carried out simultaneously with the IRAM 30-m telescope. To estimate the 
\hdos\ column density we also observed simultaneously the \uc\ line of \tco\ 
and \cdo, the \du\ line of \cdo\, and the \td\ line of CS. The half power beam 
width of the telescope was 24\as, 17\as\ and 12\as\ for the 3, 2 and 1.3 mm 
bands. The receivers, equipped with SIS mixers, were tuned to single side band 
with image rejections of \gsim 10 dB. The typical system temperatures were 
300, 500 and 900 K for the 3, 2 and 1.3 mm lines respectively. We used two filter banks of 256$\times$1MHz and one of 
512$\times$1MHz as 
spectrometers. The velocity resolution provided by the filter banks were 3, 
2 and 1.3 \kms for the 3, 2 and 1.3 mm bands respectively. The calibration was 
achieved by observing hot and cold loads. The line intensities are given in 
units of \ta*.

The observed molecular clouds were selected from the SiO maps of 
\cite{martin97} and their locations are shown in the upper panel of Fig. 1. The typical line 
profiles of \c2h5oh, \tco\ and CS towards selected clouds in SgrA (M-0.11-0.08), 
in Sgr B2 (M+0.76-0.05), and associated with the Sickle (M+0.20-0.03) and the 
NTRA (M+0.17+0.01) are shown in the lower panels of Fig. 1. Table 1 summarizes the results obtained 
for all molecular clouds. The emission in the \ct\ line of \c2h5oh is 
widespread in Sgr A and Sgr B2. However, the line profiles of \c2h5oh 
are different from those of CS, \tco, and \cdo. 
This is illustrated in Fig. 1 
for M+0.76-0.06. The CS and the \tco\ emission appears from -20 \kms to 
120 \kms. However, like the SiO emission in Sgr A and Sgr B2 \citep{martin00}, the 
\c2h5oh emission mainly appears for radial velocities 
between 10 and 70 \kms. This indicates variations in 
the abundance of \c2h5oh in the molecular clouds along the line of sight  with 
different radial velocities.  Column 4 in Table 1 gives the radial velocity of 
the peak intensity of the \ct\ \c2h5oh line and the velocity range where the 
\c2h5oh emission is  detected.

The morphology of the \c2h5oh emission in the GC is also different from those
of CS, \tco, and \cdo. \c2h5oh emission is not detected towards M-0.08-0.06 
in SgrA.
Furthermore, the \c2h5oh emission shows different behavior towards the 
different kind of filaments in the GC. In general, the molecular clouds 
associated with the thermal features like the Sickle  \citep[M+0.18-0.04 and 
M+0.20-0.03]{serabyn91} and the TAFs E1 and 
E2 \citep[M+0.13+0.02 and M+0.17+0.01]{serabyn87} do not show \c2h5oh emission. 
An interesting exception is 
the detection of \c2h5oh emission towards M+0.04+0.03, a molecular cloud 
located south of the TAF W1. However, the rather narrow radial velocity range 
of the \c2h5oh emission as compared with those of the CS and \tco \ emission 
\citep{serabyn87} suggests that the bulk of the molecular gas 
associated with thermal features does not show \c2h5oh emission. Mapping of 
the \c2h5oh emission in this region will tell the possible association of this 
velocity component with the thermal features. In contrast to the lack of 
\c2h5oh emission from the molecular clouds associated with the thermal 
features, weak \c2h5oh emission is detected towards M+0.17+0.01 in  
the NTRA.

In summary, \c2h5oh emission in the GC molecular clouds is widespread, but it 
shows substantial differences between the different features in the GC.
Furthermore, the \c2h5oh emission is mainly restricted to molecular 
clouds with radial velocities between 10 and 70 \kms.

\section{Excitation and abundance of \c2h5oh in the GC}

The lack of \c2h5oh emission towards some molecular clouds in the GC with CS 
emission cannot be due to excitation effects since the \ct\ line of \c2h5oh 
and the CS J=3-2 line have similar critical densities. The difference 
between CS and \c2h5oh emissions must be due to changes in the abundance of these molecules.
The three \c2h5oh lines can be combined to estimate the \hdos\ densities and 
the \c2h5oh column densities. The excitation temperature, \tex, derived from 
the column densities of the \ct\ and \no\ lines are given in column 5 of Table 
1. For the sources detected in both lines, the derived \tex\ ranges from 9 to 
14 K. Considering a collisional cross section of \c2h5oh similar to that of  
CH$_3$OH, the \tex\ derived from the \c2h5oh lines indicates \hdos\ densities 
of few \E{4}cm$^{-3}$. These densitites are in good agreement with those 
obtained from SiO and CS  for the GC clouds 
\citep{martin97,hutte98,serabyn87,serabyn91}.
For the other sources, the upper limits are between 8 and 13 K. In the 
following we will assume a \tex\ of 7 K for these sources.

The total \c2h5oh column densities have been derived by assuming optically 
thin emission, and using the three substates partition function given by 
 \citet{pearson97} for the \tex\ in Table 1.  Since the {\it $gauche+$} 
 and {\it gauche--} 
torsional substates are not excited, only the {\it trans} substate has been 
considered. Since the \c2h5oh profiles are, in general, 
different from those of CS and \tco, in Table 1 we give two 
different column densities for \c2h5oh. The first one derived for the 
velocity range (column 4 in Table 1) where the \c2h5oh line has been detected 
('yes' column), and the second one the upper limit derived for the velocity range with CS and 
\tco, but without \c2h5oh emission ('no' column). Table 1 also gives the 
column densities of \tco\ and CS derived for the two 
velocity ranges using the typical conditions in the GC, an 
\hdos\ density of 3 \E{4}\cmm3 and a kinetic temperature of \apro80 K.

The ratio between CS and \tco\ column densities for all sources and the two 
velocity ranges is fairly constant with a value of \apro\E{-2}. This indicates 
that the CS abundance in the GC molecular clouds is roughly constant. For 
the typical \tco\ to \hdos\ abundance ratio in the GC of 5 \E{-6}, we derive a 
CS abundance of \apro 5 \E{-8}. For this CS abundance we derive a \c2h5oh 
abundance (see column 9 of Table 1) of 0.4-5$\times$\E{-8} for the clouds with 
\c2h5oh emission. For the other clouds such as those associated with the 
thermal features, the \c2h5oh abundance decreases by at least more than one order of 
magnitude. We conclude that large \c2h5oh abundance of a few 
\E{-8} is found in most of the GC molecular clouds in the Sgr A and Sgr B2 
complexes with radial velocities between 10 and 70 \kms and the NTRA. However, 
the \c2h5oh abundance is not uniform and drops by more than one order of 
magnitude towards some molecular clouds in both complexes and the material 
associated with the thermal features in the GC.

\section {Discussion}
 
The widespread large abundance of \c2h5oh in most of the GC clouds is a clear 
signature that large scale grain mantle erosion is taking place in this region 
of the Galaxy. The abundance of \c2h5oh in the GC clouds is even larger 
than those measured  in the  hot cores in the galactic disk 
\citep{nummelin98}. In 
hot cores, the high \c2h5oh abundance is explained by grain surface chemistry  
and subsequent thermal evaporation from the grains when they are heated by 
recently formed massive stars \citep{millar91,charnley95}. In the 
GC clouds, 
most of the dust is cold with dust temperatures of 20-30 K 
\citep{martin99b,rodriguez00} which are too low to evaporate the icy mantle 
from the grains. 
Further support for the association of \c2h5oh with the cold dust comes from 
the low \c2h5oh abundance measured toward the thermal features such as the 
Sickle that show large column densities of warm dust 
\citep{simpson97}. This is likely due to the fact that 
the UV 
photons that heat the dust also photodissociate the thermal evaporated 
\c2h5oh. Thus, thermal evaporation of grain mantles in the GC clouds does not 
seem to account for the measured \c2h5oh abundance in the GC.

Shock waves are thought to dominate the heating and the chemistry of 
refractory elements in the GC \citep{wilson82,martin97,hutte98}.
\c2h5oh sputtered off the grain mantles by shock waves 
can explain the observed widespread large abundance of this molecule 
associated with the cold dust in the GC. C-shocks with moderate velocities of 
30-40 \kms can produce substantial grain processing 
\citep{flower96,caselli97,charnley00} 
explaining  the large abundance of SiO, \c2h5oh and HOCO$^+$ in 
the GC. The picture emerging from all the molecular 
data is that the chemistry in the GC is largely driven by widespread shocks 
with moderate velocities of \lsim 40 \kms. 

One interesting aspect of alcohol chemistry is its transient nature 
\citep{millar91,charnley95}. 
The typical time scale for ethanol destruction after 
the injection into gas-phase is \apro \E{4} years for the hot core 
conditions \citep{charnley95}. For the typical \hdos\ densities in the GC 
clouds, an order of magnitude smaller than those for the hot cores, the 
estimated time scale for destruction of \c2h5oh would be \apro \E{5} years. 
 This indicates that the widespread shocks that drive the chemistry in the GC 
have occurred in the last \apro \E{5} years.  Similar time scales for the 
shocks have been derived from the non-equilibrium \hdos\ ortho-to-para ratios in two GC clouds 
\citep{rodriguez00}.

Several possibilities have been proposed for the origin of the widespread shocks in the GC: 
shocks due to cloud-cloud collisions associated with the large scale dynamics 
in the context of a bar potential 
\citep{wilson82,hasegawa94,hutte98} 
and shocks produced by wind-blown bubbles driven by evolved massive stars
\citep{martin99a}. 
Both mechanisms seem to account for the time scales derived from the abundance of \c2h5oh. 
The time scale derived for the cloud-cloud collisions would be less than the galactic 
rotation period of \apro \E{6} years \citep{gusten89}. This scenario is supported by 
the large SiO abundances found at the outer inner Lindblad resonance in the 
Milky Way \citep{hutte98} and in NGC253 \citep{burillo00}. The wind-blown 
bubbles driven by evolved massive stars have typical dynamical ages 
of \apro \E{5} years, consistent with the life time derived from ethanol. 
Furthermore, evolved massive stars also explain the correlation between the 
large \c2h5oh and SiO abundance in the molecular clouds with radial 
velocities between 10 and 70 \kms and the Fe 6.4 keV lines 
\citep{martin00}. However, this scenario can only explain the 
widespread large \c2h5oh and SiO abundance if a burst of massive 
star formation has occurred in the GC \apro \E{7} years ago. 
Large scale mapping of the GC in molecular species dominated by 
different type of chemistry will help to establish the origin of the shocks.
Furthermore, the strong \c2h5oh line intensities towards many sources in the 
GC makes
these regions ideal targets to search for the large ether molecules predicted 
by \cite{charnley95}.

\acknowledgments {This work has been partially supported by 
the Spanish CICYT and the European Commission
under grant numbers ESP-1291-E and 1FD1997-1442. 
NJR-F has been supported by the Consejer{\i}a
de Educaci\'on y Cultura de la Comunidad de Madrid.}

%

\clearpage

\figcaption[fig.eps]{ Upper panel) Location of the position observed in 
\c2h5oh, \tco, \cdo\ and CS shown by filled circles superposed on the 
spatial distribution of the SiO emission from \cite{martin97}. The contour 
levels from 7.3 to 94.5 by 17.4 K \kms. Lower panel)  
Profiles of the \tco\ \uc, CS \td, and the \ct\ and \no\ 
\c2h5oh (Ethanol 4--3, Ethanol 9--8) lines towards the selected position 
shown in the upper panel.
\label{fig1}}

\begin{deluxetable}{lccrrrrrrrrrrrrr}
\tablecolumns{16}
\tablewidth{0pc}
\tablecaption{Derived physical conditions}
\rotate
\tabletypesize{\footnotesize}
\tablehead{

\colhead{Source} & \colhead{$\alpha$} & \colhead{$\delta$} & \colhead{V$_0$ (V$_i$,V$_f$)\tablenotemark{a}} & 
\colhead{T$_{ex}$\tablenotemark{b}} & \multicolumn{2}{c}{N($^{13}$CO)} &&  
\multicolumn{2}{c}{N(CS)} && \multicolumn{2}{c}{N(eth.)} && \multicolumn{2}{c}{[eth.]/[CS]}\\

\colhead{} & \colhead{17$^{\rm h}42^{\rm m}$} & \colhead{$-28\degr$} & \colhead{(km s$^{-1}$)} & \colhead{(K)} &
\multicolumn{2}{c}{(10$^{16}$\,cm$^{-2}$)} && \multicolumn{2}{c}{(10$^{14}$\,cm$^{-2}$)} && 
\multicolumn{2}{c}{(10$^{14}$\,cm$^{-2}$)} && \multicolumn{2}{c}{$10^{-1}$}\\

\cline{6-7} \cline{9-10} \cline{12-13} \cline{15-16} 

\colhead{} & \multicolumn{2}{c}{(B1950)} & \multicolumn{2}{c}{} & \colhead{yes} & \colhead{no}  && \colhead{yes} & 
\colhead{no} && \colhead{yes} & \colhead{no} && \colhead{yes} & \colhead{no} 
}
\startdata
M-0.11-0.08    & 28.0 & 62.9 &        19.8\,(5,\,32)   &      11    &           5&           3&&           
15&           2&&           7&       \lsim0.4&&    4.7&      \lsim2.2 \\
M-0.08-0.06    & \phn30.0 & 61.0 &        29.7\,(19,\,42)  &       \lsim9    &           7&           5&&           
6&           3&&           2&      \lsim0.8&&      2.8&     \lsim2.4 \\
M-0.04-0.03    & \phn29.1 & 58.1 &              (-20,\,103) &  \nodata   &  \nodata   &          15&&  
\nodata   &          15&&  \nodata   &     \lsim0.4&&  \nodata   &    \lsim0.3 \\
M-0.02-0.07    & \phn40.0 & 58.0 &        47.4\,(36,\,67)  &      14    &          44&  \nodata   &&          
22&  \nodata   &&          11&  \nodata   &&         4.7 &  \nodata   \\
M+0.04+0.03    & \phn26.2 & 51.8 &      -30.6\,(-38,\,-25) &     $\approx$ 10    &           4&          12&&           
4&           8&&           1&        \lsim0.6&&    2.8 &     \lsim0.7 \\
M+0.07-0.07    & \phn54.2 & 53.5 &        52.8\,(37,\,67)  &      12    &          10&           7&&          
12&           5&&           5&       \lsim0.6&&     3.9 &     \lsim1.1 \\
M+0.13+0.02    & \phn41.4 & 47.6 &               (-39,\,95) &  \nodata   &  \nodata   &          17&&  
\nodata   &          17&&  \nodata   &      \lsim0.5&&  \nodata   &    \lsim0.3 \\
M+0.17+0.01    & \phn50.0 & 45.8 &         59.8\,52,\,69)  &     $\approx$13    &           9&  \nodata   &&           
7&  \nodata   &&           1&  \nodata   &&     0.8 &  \nodata   \\
M+0.18-0.04    & \phn61.0 & 47.3 &                  (5,\,90)&  \nodata   &  \nodata   &          13&&  
\nodata   &          10&&  \nodata   &     \lsim0.3&&  \nodata   &    \lsim0.3 \\
M+0.20-0.03    & \phn63.6 & 45.7 &                (3,\,100) &  \nodata   &  \nodata   &          19&&  
\nodata   &          13&&  \nodata   &      \lsim0.2&&  \nodata   &     \lsim0.2 \\
M+0.24+0.01    & \phn59.6 & 42.6 &       36.4\,(23,\,53)   &      12    &          10&           8&&           
10&           4&&           8&    \lsim0.4&&     8.4 &     \lsim1.1 \\
M+0.59-0.02    & 116.4 & 25.3&       73.5\,(50,\,94)  &     \lsim11   &           6&  \nodata   &&           
4&  \nodata   &&           1&  \nodata   &&     3.2 &  \nodata   \\
M+0.62-0.10    & 137.0 & 26.5&          56.5\,(42,\,72)&      10    &           6&           2&&           
6&           1&&           3&      \lsim0.3&&      5.9 &     \lsim2.1 \\
M+0.64-0.08    & 137.5 & 24.5&        60.8\,(42,\,72)  &      \lsim9    &           5&           4&&           
6&           4&&           2&      \lsim0.5&&     3.3 &     \lsim1.3 \\
M+0.67-0.06    & 137.0 & 22.5&          50.6\,(31,\,78)&      $\approx$ 9    &          10&           3&&          
14&           2&&           3&      \lsim0.4&&    1.9 &     \lsim1.9 \\
M+0.68-0.10    & 147.2 & 23.3&        22.1\,(0,\,40)  &      $\approx$ 9    &          10&           3&&          
13&           1&&           5&      \lsim0.5&&    3.7 &    \lsim3.4 \\
M+0.70-0.01\tablenotemark{c}    & 130.0 & 19.5&        62.2\,(52,\,74)&      13    &  \nodata   &  \nodata   &&  
\nodata   &  \nodata   &&           3&  \nodata   &&  \nodata   &  \nodata   \\
M+0.70-0.09    & 147.2 & 22.1&       43.0\,(29,\,62)&     \lsim10    &           8&           8&&           
5&           5&&           2&       \lsim1.0&&       3.3&     \lsim2.2 \\
M+0.71-0.13    & 158.4 & 22.4&       42.0\,(13,\,74)  &      \lsim9    &          13&           6&&          
9&           4&&           2&       \lsim1.2&&         2.7&     \lsim3.1 \\
M+0.76-0.05    & 147.2 & 17.6&       32.4\,(14,\,59)   &     \lsim8    &          13&          10&&           
13&          10&&           6&      \lsim0.4&&         5.0 &     \lsim0.4 \\

\enddata
\tablenotetext{a}{\ V$_0$ is the LSR velocity at the peak intensity 
of the \c2h5oh \ct line. When this line is detected, (V$_i$, V$_f$) is the 
velocity range where \c2h5oh
emission is detected. When \c2h5oh is not detected, (V$_i$, V$_f$) represent 
the full velocity range of the CS and \tco\ emission.}

\tablenotetext{b}{\ T$_{ex}$ is the excitation temperature derived from 
the \ct\ and the \no lines of \c2h5oh (see text).}

%
\end{deluxetable}

\begin{thebibliography}{}

\bibitem[Caselli, Hartquist and Havnes(1997)]{caselli97} 
Caselli, P., Hartquist, T.\ W.\ and Havnes, O.\ 1997, \aap, 322, 296 

\bibitem[Charnley, Tielens and Millar(1992)]{charnley92} Charnley, S. B., 
Tielens, A. G. G. M. and Millar, T. J., 1992, \apjl, 399, L71--L74

\bibitem[Charnley et al.(1995)]{charnley95}
Charnley, S. B., Kress, M. E., Tielens, A. G. G. M. and Millar, T. J., 1995
\apj, 448, 232

\bibitem[Charnley and Kaufman(2000)]{charnley00}
Charnley, S. B. and Kaufman, M. J., 2000, \apjl, 529, L111--L114

\bibitem[Flower et al.(1996)]{flower96} 
Flower, D.\ R., Pineau des F\^orets, G., Field, 
D.\ and May, P.\ W.\ 1996, \mnras, 280, 447

\bibitem[Garc{{\i}}a-Burillo, et al.(2000)]{burillo00} 
Garc{{\i}}a-Burillo, S., Mart{{\i}}n-Pintado, 
J., Fuente, A.\ and Neri, R.\ 2000, \aap, 355, 499 

\bibitem[G{\"u}sten (1989)]{gusten89} G{\"u}sten, R.\ 1989, IAU 
Symp.\ 136: The Center of the Galaxy, 136, 89 

\bibitem[Hasegawa et al.(1994)]{hasegawa94} 
Hasegawa, T., Sato, F., Whiteoak, J.\ B.\ and Miyawaki, R.\ 1994, \apjl, 
429, L77 

\bibitem[H\"uttemeister et al.(1993)]{hutte93}  H\"uttemeister, S., Wilson, T. 
L., Bania, T. M. and Mart{\i}n-Pintado, J. 1993, \aap, 280, 255


\bibitem[H\"uttemeister et al.(1998)]{hutte98} Huettemeister, 
S., Dahmen, G., Mauersberger, R., Henkel, C., Wilson, T.\ L.\ and 
Mart{\i}n-Pintado, J.\ 1998, \aap, 334, 646 

 
\bibitem[Irvine, Goldsmith and Hjalmarson(1987)]{irvine87}
Irvine, W. M., Goldsmith and P. F., Hjalmarson, A., 1987, 
Interstellar Processes; Proceedings of the Symposium, D. Reidel 
Publishing Co., 561--609.



\bibitem[Lang, Morris and Echevarria(1999)]{lang99} Lang, C.\ 
C., Morris, M.\ and Echevarria, L.\ 1999, \apj, 526, 727


\bibitem[Mart{\i}n-Pintado et al.(1997)]{martin97}  Mart{\i}n-Pintado, J., de 
Vicente, P., Fuente, A., and Planesas, P. 1997, ApJ, 482, L45

\bibitem[Mart{\i}n-Pintado et al.(1999a)]{martin99a} Mart{\i}n-Pintado, J., 
Gaume, R. A., Rodr{\i}guez-Fern\'andez,
 N., De Vicente P., and Wilson, T. L. 1999a, ApJ, 519, 667

\bibitem[Mart{\i}n-Pintado et al.(1999b)]{martin99b} 
Mart{\i}n-Pintado, J., Rodriguez-Fernandez, N.\ J., de Vicente, P., Fuente, 
A., Wilson, T.\ L., Huttenmeister, S.\ and Kunze, D.\ 1999b, ESA SP-427: The 
Universe as Seen by ISO, 427, 711 

\bibitem[Mart{\i}n-Pintado et al.(2000)]{martin00}
Mart{{\i}}n-Pintado, J., de Vicente, P., Rodr{{\i}}guez-Fern{\'a}ndez, N.\ J. ,
Fuente, A. and Planesas, P., 2000, \aap, 356, L5--L8

\bibitem[Millar et al.(1988)]{millar88}
Millar, T. J., Brown, P. D., Olofsson H. and Hjalmarson H., A., 1988,
\aap, 205, L5--L7

\bibitem[Millar, Herbst, and Charnley(1991)]{millar91}
Millar, T. J., Herbst, E. and Charnley, S. B., 1991, \apj, 369, 147--156

\bibitem[Millar, MacDonald and Habing(1995)]{millar95} Millar, T. J., MacDonald
G. H. and Habing, R. J., \mnras, 1995, 273, 25--29

\bibitem[Minh, Irvine and Friberg(1992)]{minh92}
Minh, Y. C., Irvine, W. M. and Friberg, P., 1992, \aap, 258, 489--494.

\bibitem[Nummelin et al.(1998)]{nummelin98} Nummelin, A., Dickens, J. E., 
Bergman, P., Hjalmarson, A., Irvine, W. M., Ikeda, M. and Ohishi, M., 
1998, \aap, 275--286

\bibitem[Pearson et al.(1997)]{pearson97} 
Pearson, J.\ C., Sastry, K.\ V.\ L.\ N., Herbst, E.\ and de Lucia, F.\ C.\ 
1997, \apj, 480, 420

\bibitem[Ohishi et al.(1995)]{ohishi95} Ohishi, M., Ishikawa, 
S., Yamamoto, S., Saito and S., Amano, T., 1995 \apjl, 446, L43

\bibitem[Rodr{\i}guez-Fern{\'a}ndez et al.(2000a)]{rodriguez00} 
Rodr{\i}guez-Fern{\'a}ndez, N.\ J., Mart{{\i}}n-Pintado, J., de Vicente, 
P., Fuente, A., H{\"u}ttemeister, S., Wilson, T.\ L.\ and Kunze, D.\ 2000, 
\aap, 356, 695 

\bibitem[Rodr{\i}guez-Fern{\'a}ndez et al.(2001)]{rodriguez01}
     Rodr{\i}guez-Fern\'andez N.J., Mart{\i}n-Pintado J., Fuente A., de
Vicente P.,
     Wilson T.L., H\"uttemeister S., 2001, A$\&$A  in press


\bibitem[Serabyn and G\"usten(1987)]{serabyn87} Serabyn, E.\ and 
G\"usten, R.\ 1987, \aap, 184, 133 
\bibitem[Serabyn and G\"usten(1991)]{serabyn91} Serabyn, E.\ and 
G\"usten, R.\ 1991, \aap, 242, 376


\bibitem[Simpson et al.\ (1997)]{simpson97} Simpson, J.\ P., 
Colgan, S.\ W.\ J., Cotera, A.\ S., Erickson, E.\ F., Haas, M.\ R., Morris, 
M.\ and Rubin, R.\ H.\ 1997, \apj, 487, 689

\bibitem [Wilson et al.(1982)]{wilson82}
Wilson, T. L., Ruf, K., Walmsley, C. M., Martin, R. N., Batrla, W. and Pauls, 
T. A. 1982 \aap, 115, 185

\bibitem[Zuckerman et al.(1975)]{zuckerman75} Zuckerman, B., Turner, 
B. E., Johnson, D. R., Lovas, F. J. Fourikis, 
N., Palmer, P., Morris, M., Lilley, A. E., Ball, J. A. and Clark, F. O., 1975
\apjl, 196, L99--L102

\end{thebibliography}
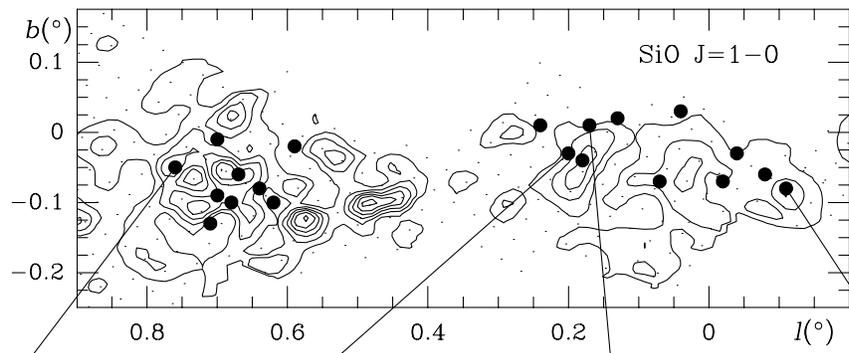
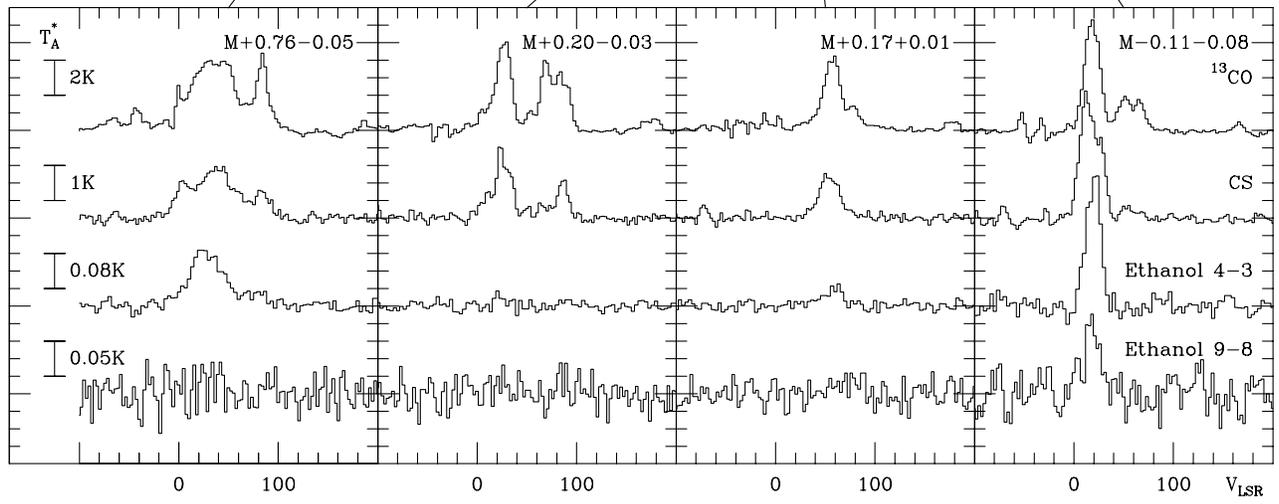
\end{document}